\documentclass[preprint]{revtex4}

\usepackage{graphicx}
\usepackage{amssymb}
\usepackage{longtable}

\begin{document}

\preprint{PREPRINT}

\title{Thermodynamic properties of short-range attractive Yukawa fluid: Simulation and theory}

\author{Pedro Orea}
\affiliation{Programa de Ingenier\'{\i}a Molecular, Instituto
Mexicano del Petr\'{o}leo, Eje Central L\'{a}zaro C\'{a}rdenas
152, 07730 M\'{e}xico D.F., M\'exico.}

\author{Carlos Tapia-Medina}
\affiliation{Departamento de Energ\'{\i}a, Universidad Aut\'onoma
Metropolitana-Azcapotzalco, Av. San Pablo 180, Col. Reynosa, 02200
M\'exico D.F., M\'exico.}

\author{Davide Pini}
\affiliation{Dipartimento di Fisica, Universit\`a degli Studi di
Milano, Via Celoria 16, 20133 Milano, Italy}

\author{Albert Reiner}
\affiliation{Erzbisch\"ofliches Priesterseminar, Boltzmanngasse 7-9/303, A-1090
    Vienna, Austria}

\begin{abstract}

Coexistence properties of the hard-core attractive Yukawa
potential with inverse-range parameter $\kappa = 9, 10, 12$ and
$15$ are calculated by applying canonical Monte Carlo simulation.
As previously shown for longer ranges, we show that also for the
ranges considered here the coexistence curves scaled by the
critical density and temperature obey the law of corresponding
states, and that a linear relationship between the critical
density and the reciprocal of the critical temperature holds. The
simulation results are compared with the predictions of the
self-consistent Ornstein-Zernike approximation, and a good
agreement is found for both the critical points and the
coexistence curves, although some slight discrepancies
are present.

\end{abstract}

\maketitle

\section{Introduction}

Short-range potential models of simple analytic form
have been the object of intense investigation over the last years.
The main reason for such an interest is that they provide an
approximate representation of the effective interactions and phase
behavior experimentally observed in a variety of real systems. For
instance, globular protein solutions, colloidal suspensions, and
fullerenes at high temperature~\cite{foffi02,prausnitz04,caccamo00,
fuchs99,zukos99}. This scenario has been further enriched by
recent investigations of gas-solid, solid-solid,  and a metastable
vapor-liquid transition in one of the most studied short-range
models, namely the hard-core attractive Yukawa (HCAY)
potential~\cite{frenkel94,dijkstra02,fu03,zhou04,tuinier06,reichman07,germain07}.
The HCAY model is given by

\begin{equation}
\label{pot}
u(r)=\left\{\begin{array}{ll}\infty, & \mbox{ if $r<\sigma,$}\\
             -\epsilon \sigma \exp[-\kappa (r-\sigma)]/r, & \mbox{ if $ \sigma\leq
             r, $}
             \end{array} \right.
\end{equation}

\noindent where $\epsilon$ is the depth of the potential and
$\sigma$ is the hard-sphere diameter. The range of the potential 
is tuned by varying
$\kappa$. Our results are given in dimensionless units, such that
$r^*=r/\sigma$ for distance, $T^*=k_{B}T/\epsilon$ for temperature
($k_B$ is the Boltzmann's constant), $\rho^*=\rho\sigma^3$ for
density.

Many theoretical approaches have been used for studying the HCAY
fluid, for example: the Barker-Henderson perturbation
theory~\cite{prausnitz04,paricaud06}, the thermodynamic
perturbation theory
(TPT)~\cite{zhou04,zhou06,haro07,duglas09,singhFPE09} and a variety 
of other mean field approaches based on truncation 
or approximate summation of perturbative series~\cite{weiss07,weiss09}, 
the mean spherical approximation
(MSA)~\cite{blum95,herrera96,mier97,tuinier06}, the modified hypernetted 
chain approximation (MHNC)~\cite{caccamo99} and other integral equation
theories~\cite{jakse06,ayadim09}, the density
functional theory~\cite{fu03,fu04}, 
the self-consistent Ornstein-Zernike approximation
(SCOZA)~\cite{hoye98,pini98,foffi02,hoye08,weis07}, 
and the hierarchical reference theory (HRT)~\cite{caccamo99,pini09}, 
among others. For
all theories, the most difficult regime to deal with is that of
short interaction range, where an accurate localization of the
vapor-liquid coexistence curve and of the critical point
presents severe difficulties. Sometimes, the shape of
the coexistence curve is not reproduced correctly and critical
points are not located correctly either, especially the critical
density. We found that the critical density data reported by these
approaches show different tendencies for shorter-range
attractions, when we plot $\rho^*_c$ versus $1/T^*_c$. These
issues were some of the motivations to carry out this
work.

For large values of $\kappa>7$, the computer simulation data for
HCAY fluid are quite scarce. In this regime, Gibbs ensemble Monte
Carlo (GEMC) cannot reproduce binodal
curves~\cite{frenkel94,lomba94}. The phase diagram at large
$\kappa$ was studied by Monte Carlo supplemented by thermodynamic
integration (MC-TI) in Ref.~\cite{dijkstra02}, but the results do
not include the critical density. In our previous works we
reported the coexistence properties of HCAY fluid for long range
tails. These properties were compared to those reported in the
literature, and some differences were found among
them~\cite{yuka1,yuka2,yuka3}. In a recent work,
Singh~\cite{singh09} has reported the coexistence curves of
short-range attractive Yukawa (SR-HCAY) fluid with $\kappa=8-10$,
using Grand-canonical transition-matrix Monte Carlo (GC-TMMC) with
the histogram reweighting method.
However, this method   
proves difficult to use in order 
to calculate coexistence properties at low
temperature and high liquid densities~\cite{singh09}. So, NVT-MC 
simulations appear to be a more efficient method to calculate such properties
for very short-range systems, where these conditions are met.

In view of the above considerations, the main goals of this paper
are to demonstrate that there is a linear dependence between the
critical density and the reciprocal of the critical temperature
even for short interaction ranges, where theoretical approaches
show different tendencies, and to report a systematic study of the
liquid-vapor phase diagrams of SR-HCAY fluid with $\kappa=9, 10,
12$ and $15$, using canonical ensemble Monte Carlo
simulation, where most theoretical
approaches fail and other simulation techniques cannot reproduce
the binodal curves. Finally, once again we have confirmed that the
coexistence curves of SR-HCAY model follow the law of corresponding
states, i.e., curves corresponding to different $\kappa$ fall on top
of each other within a high degree of accuracy, provided  
the density $\rho$ and temperature $T$ are rescaled by their critical
values $\rho_{c}$, $T_{c}$.  

Besides, we have also reported the binodal curves and critical
points predicted by SCOZA for the above values of $\kappa$.
Although the application of SCOZA to narrow Yukawa potentials has
been considered before~\cite{caccamo99,foffi02}, an extensive
comparison with simulation data for the phase diagram in this
regime has not been possible so far because of the aforementioned
paucity of simulations at large $\kappa$, and the present study
represents a good opportunity to perform it. Results were
obtained both by the standard version of SCOZA considered in
Refs.~\cite{pini98,caccamo99,foffi02}, and by a modified version
developed in~\cite{hoye08}, where consistency with the virial
route, which is disregarded in the original implementation, is
partially taken into account. The comparison shows that some
discrepancies between SCOZA and simulations are definitely
present: specifically, the standard formulation of SCOZA
overestimates both the critical density and critical temperature,
while the converse happens with the modified formulation. The
latter also predicts a nearly constant critical density at values of $\kappa$,
for which the simulation show that this quantity still changes significantly. 
Nevertheless, the overall agreement with the simulations is quite
satisfactory.

The paper is organized as follows: in Sec.~\ref{sec:simu} the details of our
simulations are briefly summarized; in Sec.~\ref{sec:scoza}, an overview
of both the standard and the modified versions of SCOZA is given, and some
issues relevant for the application of the theory to short-range interactions
are discussed; in Sec.~\ref{sec:results} our results are presented and
discussed; in Sec.~\ref{sec:conclusions} our conclusions are drawn.

\section{Simulation Details}
\label{sec:simu}

Canonical ensemble Monte Carlo (NVT-MC) simulations of HCAY
fluid interface have been performed using Metropolis algorithm. To
prepare the initial configuration we placed $N = 1726$ particles
in a face-centered cubic array in the middle of the simulation
cell, which allowed us to obtain two interfaces with a vapor phase
surrounding the liquid when the system is equilibrated (see Fig 2
of Ref.\cite{yuka1}).\\

Simulations were performed on a parallelepiped cell with
dimensions $L_x=L_y=10$, and $L_z=40$. Additional MC runs were
carried out with $L_x=L_y=14$ and a larger number of
particles~\cite{jll05,malfreyt09} to check for finite-size
effects, and no such effect was found. The cutoff radius was
selected to $R_c=2$. Periodic boundary conditions were applied in
all three directions and the neighbor list was applied to speed up
calculation. The maximum particle displacement was adjusted to
give a $45\%$ acceptance rate. The simulations were performed in
cycles; in every cycle it was attempted to move all the particles
that are in the neighbor list. The system was equilibrated for
$10^6$ cycles and the coexistence properties were calculated for
$10^7$ cycles divided into $100$ blocks of $10^5$ cycles each one,
in order to avoid density fluctuation.\\

The density profiles, $\rho(z)$, were calculated every 50 cycles
and at the end of each simulation run, every profile was fitted to
a hyperbolic tangent function to obtain $\rho^*_V$ and $\rho^*_L$
values\cite{yuka1,yuka2,yuka3}. The critical parameters for the
HCAY fluid were calculated by using the rectilinear diameter
law~\cite{landau00} and the universal value $\beta_{\it coex}=0.325$ 
for the critical exponent $\beta_{\it coex}$ which gives the curvature 
of the liquid-vapor coexistence curve in the critical region. All
the critical point parameters we calculated are listed in Table~\ref{table1}.

\section{Theory}
\label{sec:scoza}

A detailed description of SCOZA and its application to a Yukawa fluid has been
given elsewhere and we refer the interested reader to the previous literature
on this subject~\cite{hoye98,pini98,caccamo99,foffi02}. Here we recall that, 
as usual
in integral-equation theories, in the SCOZA the Ornstein-Zernike (OZ)
equation is closed by an approximate {\em ansatz} which involves the direct
correlation function $c(r)$ and the radial distribution function $g(r)$.
In the standard formulation of SCOZA, one requires that i) the core condition
$g(r)=0$ be satisfied inside the hard core $r<\sigma$; and ii)
the contribution to $c(r)$ outside the hard core due to tail of the potential
be linear in the potential itself. For a HCAY potential one has then:
\begin{equation}
\left\{
\begin{array}{ll}
g(r)=0                     & \mbox{if $r<\sigma$} \, , \\
c(r)=c_{\it HS}(r)+K \exp[-\kappa (r-\sigma)]/r
\mbox{\hspace{0.2cm}} & \mbox{if $\sigma\leq r$}  \, ,
\end{array}
\right.
\label{closure}
\end{equation}
where $c_{\it HS}(r)$ is the direct correlation function of the hard-sphere
fluid. The functional form of this closure is similar to that
of well-established theories such as the mean spherical approximation (MSA),
where $c_{\it HS}(r)$ is identically vanishing for $r>\sigma$,
or the optimized random-phase approximation (ORPA)~\cite{hansen}. However,
in these approaches the amplitude $K$ of the direct correlation function
coincides with the inverse temperature $\beta=1/(k_{B}T)$, while in the SCOZA
$K$ is an {\em a priori} unknown function of the thermodynamic
state, to be determined so that consistency between the compressibility and
the internal energy route to the thermodynamic is obtained. The consistency
requirement is embodied in the following condition on the reduced
compressibility $\chi_{\it red}$ and the excess internal energy per unit
volume $u$:
\begin{equation}
\frac{\partial}{\partial \beta}\left(\frac{1}{\chi_{\it red}}\right)=
\rho \frac{\partial^2 u}{\partial \rho^2} \, ,
\label{consist}
\end{equation}
where $\chi_{\it red}$ and $u$ are obtained via the compressibility
and internal energy route respectively. Eq.~(\ref{consist}) together with
closure~(\ref{closure}) give a closed partial differential equation (PDE)
for $u$, that can be solved numerically, provided the initial condition
at $\beta=0$ and the boundary conditions at low and high density are
specified. The solution procedure is made easier by the fact that,
for a given $K$, the OZ equation with closure~(\ref{closure}) can be solved
analytically, if $c(r)$ for $r>\sigma$ is given by a superposition of Yukawa
tails~\cite{hoye76}. This is indeed the case, provided one adopts the Waisman
parametrization for the hard-sphere direct correlation function
$c_{\it HS}(r)$~\cite{waisman}. According to this, $c_{\it HS}(r)$
for $r>\sigma$
is also given by a Yukawa tail with known, density-dependent amplitude
and range, so that $c(r)$ for $r>\sigma$ consists of the superposition
of two Yukawas. On the other hand, for non-Yukawa intermolecular potentials
the SCOZA $c(r)$ for $\sigma\leq r$ is not of Yukawa form, irrespective
of the parametrization adopted for $c_{\it HS}(r)$, and in order to implement
SCOZA a fully numerical solution
of the OZ equation is usually required~\cite{paschinger05,betancourt08}.

In conventional SCOZA, the virial route is not included
in the consistency
condition. In order to enforce consistency among all the three routes, one
has to endow Eq.~(\ref{closure}) with one more degree of freedom. To this
purpose, the following form for $c(r)$ was proposed~\cite{hoye08}:
\begin{eqnarray}
c(r)& \! =\! & c_{\it HS}(r)+\beta\exp[-\kappa(r-\sigma)]/r \nonumber \\ 
& & \mbox{}+H \exp[-z(r-\sigma)]/r \mbox{\hspace{0.8cm}if $\sigma\leq r$} \, ,
\label{gmsa}
\end{eqnarray}
where the amplitude of the (attractive) Yukawa potential has now been set to
$\beta$ as in the MSA, while both the amplitude $H$ and the inverse range $z$
of the extra Yukawa tail must be determined by imposing consistency among
virial, compressibility, and internal energy. This approach is very
similar to that originally proposed in Ref.~\cite{hoye84}, which is obtained 
from Eq.~(\ref{gmsa}) by having the last Yukawa to account also for the 
hard-sphere contribution to $c(r)$ outside the repulsive core, thereby setting
$c_{\it HS}(r)=0$ in Eq.~(\ref{gmsa}).
A fully numerical implementation of that closure was considered  
in Ref.~\cite{caccamo99}, but
the numerical algorithm failed to converge in the critical region. In a recent
work~\cite{hoye08}, a different strategy was adopted: on the one hand,
the analytical solution of the OZ equation for a $c(r)$ given by
a superposition of Yukawas (three when also $c_{\it HS}(r)$ is taken into
account) for $\sigma\leq r$ was again exploited. On the other hand, consistency
with the virial route was required to hold only at the critical point,
and $z$ was fixed at the value at which this condition is satisfied, thereby
making the theory considerably easier to implement. This is the approach used
here together with the standard SCOZA of Eqs.~(\ref{closure}),
(\ref{consist}).  In the following the standard and modified versions of SCOZA
will be referred to as S-SCOZA and M-SCOZA respectively.

Before concluding this Section, it is worthwhile adding some remarks
about the
relevance of the high-density boundary condition for the SCOZA PDE. While the
boundary condition at low density is trivial, since one must have
$\chi_{\it red}(\rho\!=\!0,\beta)=1$, $u(\rho\!=\!0,\beta)=0$ for every
$\beta$, the behavior at high density poses more problems. In fact, this is
not known beforehand, so that there is a certain amount of arbitrariness
on the form of the corresponding boundary condition.
For instance, in Refs.~\cite{pini98,caccamo99,foffi02},
the boundary condition was imposed on $\partial^{2}u/\partial\rho^{2}$, which
was set to the value obtained
by the so-called high-temperature approximation (HTA)~\cite{hansen}, i.e.,
by replacing $u$ with its zeroth-order contribution in an expansion in powers
of $\beta$.
In Ref.~\cite{hoye08}, instead, the boundary condition was imposed
directly on $u$ rather than on its derivative, and $u$ was determined using
the $g(r)$ obtained via the ORPA, which amounts to setting $K=\beta$
in Eq.~(\ref{closure}). 
Since there is not any obvious recipe for
singling out a specific high-density boundary condition, a natural requirement
is that the results should not depend on the detailed form of such 
a condition, or on the density $\rho_{0}$ at which it is imposed. 
This turns out to be indeed the case,
provided $\rho_{0}$ is high enough. For Lennard-Jones like tails, such as that
corresponding to the widely adopted value $z=1.8$~\cite{pini98}, setting
the high-density boundary at $\rho_{0}^*=1$ proves to be amply
sufficient. However, for interactions of shorter range, such as those
considered
in Ref.~\cite{foffi02} as well as in the present paper, $\rho_{0}$ must be
shifted to substantially higher densities. This was also observed
in a fully numerical implementation of SCOZA for a square-well
fluid~\cite{paschinger05}.
In fact, for the largest value of $z$
considered here, $z=25$ (see Table~\ref{table1}), $\rho_{0}$ must be at least
as high
as the close-packing density $\rho_{0}^*=\sqrt{2}$, and at even larger
$z$ one needs to move $\rho_{0}$ beyond close packing. This requirement sounds
quite unphysical, but one should keep in mind that here the effect of the
singular repulsion is described at the level of the Carnahan-Starling equation
of state~\cite{hansen}, which is itself unaware of close packing, and has the
compressibility vanish at the unphysical value $\rho^*=6/\pi$
rather than at $\rho^*=\sqrt{2}$. The HTA for $u$ is expected to become 
exact at close packing~\cite{stell83}, but in view of the above, it is not
surprising that one might need to go beyond close packing for the hard-sphere
repulsion to overwhelm the attractive part of the interaction in $g(r)$.

If the boundary is not located at sufficiently high $\rho_{0}$, the results
turn out to be very sensitive with respect to both a shift in $\rho_{0}$, and
the details of the boundary condition. This observation is relevant for some
of the results reported in Ref.~\cite{hoye08}, where both S-SCOZA and M-SCOZA
calculations were performed for different values of $z$, but the high-density
boundary was always kept fixed at $\rho_{0}^*=1$. This is adequate
for $z\lesssim 8$, but for larger $z$ it leads to an underestimation of both
the critical density and the critical temperature with respect to the genuine
SCOZA result, which becomes more and more severe as $z$ is increased.
The results presented here are not affected by this spurious effect, and
are meant to supersede those shown in Table~I of Ref.~\cite{hoye08}.

\section{Results and discussions}
\label{sec:results}

\begin{figure}
\includegraphics[width=8cm]{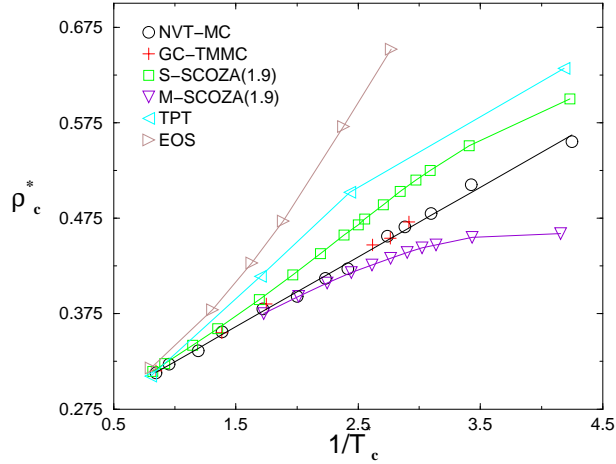}
\caption{Relation between critical density and
inverse critical temperature of HCAY model for different
simulation and theoretical results. NVT-MC with $1.8\le\kappa\le
15$ from this and previous works~\cite{yuka1,yuka2,yuka3}. GC-TMMC
with $1.8\le\kappa\le 10$~\cite{singh09}. SCOZA with
$1.8\le\kappa\le 25$ from this 
and previous works~\cite{pini98,caccamo99,foffi02},
M-SCOZA with $4\le\kappa\le 25$ from this work, EOS with
$1.8\le\kappa\le 12$~\cite{mier97}, and TPT with $1.8\le\kappa\le
25$~\cite{zhou06}.}
\label{fig:trho}
\end{figure}

We start our discussion with an analysis of the critical values of
HCAY fluid. Table~\ref{table1} presents the reduced critical
temperature $T^{*}_{c}$ and density $\rho^{*}_{c}$ for a wide
range of $\kappa$ predicted by both simulations and several
theoretical approaches, namely, the S-SCOZA and M-SCOZA described
in the previous section, the TPT by Zhou~\cite{zhou06}, and the
equation of state (EOS) by Duh and Mier-Y-Teran~\cite{mier97}.
We recall that, as $\lambda$ is increased, the liquid-vapor critical point becomes
metastable with respect to freezing. According to many theoretical and simulation 
studies~\cite{frenkel94,foffi02,dijkstra02,fu03,tuinier06,reichman07},
for $\lambda\geq 7$ the metastable regime has alreday set in.  
\begin{longtable}{ccccccccc}
\hline\hline $\hspace{0.5cm}\kappa$\hspace{0.5cm} &
\hspace{0.5cm}$T_c^*$\hspace{0.5cm}  &
\hspace{0.5cm}$\rho_c^*$\hspace{0.5cm} & \hspace{0.5cm}Method  \hspace{0.5cm} & \hspace{0.5cm} Ref. \hspace{0.5cm} \\
\hline

 1.8    & 1.180   & 0.313  & NVT-MC & \cite{yuka2} \\
 1.8    & 1.180   & 0.315  & GC-TMMC & \cite{singh09} \\
 1.8    & 1.212   & 0.312  & GC-FSS\mbox{$^a$} & \cite{pini98} \\
 1.8    & 1.219  & 0.314 & S-SCOZA &\cite{pini98} \\
 1.8    & 1.246   & 0.310 & TPT  &\cite{zhou06} \\
 1.8    & 1.240   & 0.318  & MSA-EOS & \cite{mier97} \\
 2.0    & 1.050   & 0.322  & NVT-MC &\cite{yuka2} \\
 2.0    & 1.088  & 0.323 & S-SCOZA & this work \\
 2.5    & 0.840   & 0.336  & NVT-MC & \cite{yuka2} \\
 2.5    & 0.871  & 0.342 & S-SCOZA & this work \\
 3.0    & 0.721   & 0.356  & NVT-MC &  \cite{yuka2} \\
 3.0    & 0.722   & 0.355  & GC-TMMC & \cite{singh09}\\
 3.0    & 0.740  & 0.359 & S-SCOZA & this work \\
 3.0    & 0.764   & 0.379 & MSA-EOS & \cite{mier97}\\
 4.0    & 0.581  & 0.380  &  NVT-MC &\cite{yuka1}   \\
 4.0    & 0.572  & 0.385  &  GC-TMMC & \cite{singh09} \\
 4.0    & 0.591 & 0.389 &  S-SCOZA & \cite{caccamo99}  \\
 4.0    & 0.579 & 0.375  & M-SCOZA & this work \\
 4.0    & 0.588  & 0.414  & TPT & \cite{zhou06}   \\
 4.0    & 0.614  & 0.428  & MSA-EOS & \cite{mier97}  \\
 5.0    & 0.500   & 0.39  & NVT-MC & \cite{yuka1} \\
 5.0    & 0.509  & 0.416 & S-SCOZA & this work\mbox{$^b$} \\
 5.0    & 0.497  & 0.393  & M-SCOZA & this work \\
 5.0    & 0.530   & 0.472  & MSA-EOS & \cite{mier97}  \\
 6.0    & 0.448   & 0.412  & NVT-MC & \cite{yuka1}  \\
 6.0    & 0.456  & 0.438 & S-SCOZA & this work\mbox{$^b$}  \\
 6.0    & 0.445  & 0.407  & M-SCOZA & this work \\
 7.0    & 0.414   & 0.422  & NVT-MC & \cite{yuka1} \\
 7.0    & 0.419  & 0.457 & S-SCOZA & \cite{caccamo99} \\
 7.0    & 0.409  & 0.418  & M-SCOZA & this work \\
 7.0    & 0.411   & 0.502  & TPT & \cite{zhou06} \\
 8.0    & 0.382  & 0.447  & GC-TMMC & \cite{singh09} \\
 8.0    & 0.392 & 0.474 & S-SCOZA & this work \\
 8.0    & 0.383 & 0.426  & M-SCOZA & this work \\
 9.0    & 0.365   &  0.456  & NVT-MC & this work  \\
 9.0    & 0.362   &  0.454  & GC-TMMC & \cite{singh09} \\
 9.0    & 0.370  &  0.489 & S-SCOZA & this work  \\
 9.0    & 0.368  &  0.433  & M-SCOZA & this work \\
10.0    & 0.347   &  0.466  & NVT-MC & this work  \\
10.0    & 0.343   &  0.471  & GC-TMMC & \cite{singh09} \\
10.0    & 0.352  &  0.503 & S-SCOZA & this work \\
10.0    & 0.345  &  0.439  & M-SCOZA & this work \\
10.0    & 0.361  &  0.652  & MSA-EOS & \cite{mier97} \\
11.0    & 0.337    & 0.515 & S-SCOZA & this work  \\
11.0    & 0.331    & 0.444 & M-SCOZA & this work \\
12.0    & 0.323   & 0.480 &  NVT-MC & this work  \\
12.0    & 0.324   & 0.525 & S-SCOZA & this work  \\
12.0    & 0.319   & 0.447 & M-SCOZA & this work \\
15.0    & 0.292   & 0.515 & NVT-MC & this work \\
15.0    & 0.293  & 0.551 & S-SCOZA & this work  \\
15.0    & 0.291  & 0.455 & M-SCOZA & this work \\
25.0    & 0.235  & 0.555 & MC-TI & \cite{dijkstra02}\mbox{$^c$} \\
25.0    & 0.236  & 0.600 & S-SCOZA & this work \\
25.0    & 0.241   & 0.459 & M-SCOZA & this work \\
25.0    & 0.239  & 0.632 & TPT & \cite{zhou06}  \\
\caption{Critical parameters of HCAY Fluid. \\ $^a$Gran canonical
Monte Carlo with finite-size scaling. \\ $^b$The calculation for this
value of $\kappa$ had already been performed in Ref.~\cite{foffi02}, 
but the result had not been tabulated. \\ $^c\rho_c$
was obtained using Eq.~(4) from Ref.~\cite{yuka3}.}
\label{table1}
\end{longtable}
The critical density
as a function of the inverse of the critical temperature as obtained
from our simulations and the aforementioned theories is shown
in Fig.~\ref{fig:trho}.
The critical simulation data
obtained by Singh using GC-TMMC, also displayed in the figure, are
in excellent agreement with ours, which shows that both techniques
give similar results within the error bar.
At small $\kappa$, all theories give similar results, and are in
good agreement with the simulations. In fact, in the long-range
regime even a simple mean-field approach along the lines of 
the van der Waals theory provides an accurate description of the
system, which has been shown to become exact in the
limit where the range of the tail potential goes to infinity, and its 
strength goes to zero~\cite{kac63}. As the interaction range gets shorter, discrepancies
become more and more pronounced, especially as far as
$\rho^{*}_{c}$ is concerned. In particular, the data for
$\rho^{*}_{c}$ reported by Duh and Mier-Y-Teran~\cite{mier97}
using their equation of state (EOS) for HCAY fluid strongly
overestimate the simulations. The TPT by Zhou~\cite{zhou06} and the
S-SCOZA show a better agreement with the simulation data. The
latter are bracketed by the S-SCOZA and M-SCOZA results: S-SCOZA
overestimates both $\rho^{*}_{c}$ and $T^{*}_{c}$, while the
converse is true for M-SCOZA. For both approaches, the relative
error with respect to the simulations is in general considerably
smaller for $T^{*}_{c}$ than for $\rho^{*}_{c}$.

The most remarkable feature of the simulation results is the
linear trend for $\rho^{*}_{c}$ vs. $1/T^{*}_{c}$ up to
inverse-range parameters as large as $\kappa=15$, corresponding to
$1/T^{*}_{c}=3.425$. A similar behavior is displayed by S-SCOZA,
as well as by TPT. As $\kappa$ is further increased, the slopes of
both SCOZA and TPT plots start to decrease, and deviations from
linearity become apparent. This is consistent with the fact that,
as the potential range goes to zero and $1/T^{*}_{c}$ diverges,
one expects $\rho^{*}_{c}$ to tend to a finite, non-vanishing
value. Such a qualitative behavior is predicted by Baxter's
analytical solution of his adhesive hard-sphere (AHS)
model~\cite{baxter68}, and has recently been supported by MC
simulations of the AHS model itself~\cite{miller2003,miller2004}
as well as of square-well fluids with very short attraction
range~\cite{largo2008}. It may also be recalled that for the HCAY
fluid, a finite value of $\rho^{*}_{c}$ in the limit of vanishing
attraction range is found analytically also within
MSA~\cite{mier89}. We did not perform simulations for
$\kappa>15$, but we took the critical temperature for $\kappa=25$
obtained in Ref.~\cite{dijkstra02}, and inserted it into Eq.~(4)
of Ref.~\cite{yuka3} to obtain the critical density. The result,
also plotted in Fig.~\ref{fig:trho}, shows a trend
similar to that of S-SCOZA. In this respect, it may be observed
that, although M-SCOZA has the best agreement with simulations for
$\kappa<15$, it overemphasizes the tendency to saturation at
larger $\kappa$~\cite{note}, which is better described by S-SCOZA.

\begin{figure}
\includegraphics[width=8cm]{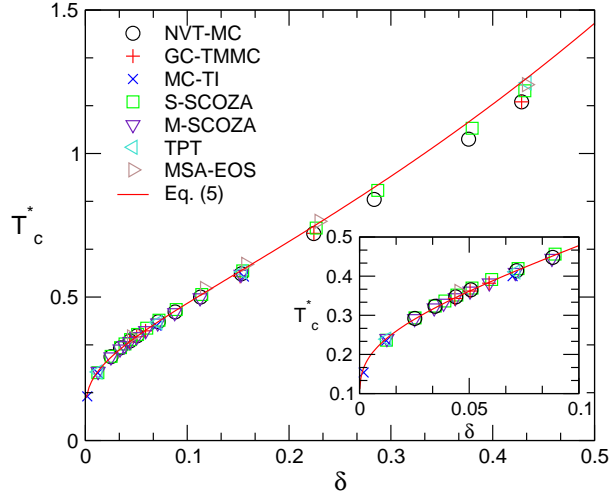}
\caption{Relation between critical temperature and
effective range $\delta$ (see text) of the HCAY model for different simulation 
and theoretical results. Symbols have the same meaning 
as in Fig.~\ref{fig:trho}. MC-TI with $3.9\leq \kappa \leq 100$ 
from Ref.~\cite{dijkstra02}.
The red line represents the prediction of Eq.~(\ref{square}). The inset
is an enlargement of the short-range region.}
\label{fig:tkappa}
\end{figure}

We now turn to the dependence of the critical temperature on the
interaction range. For hard-core plus attractive tail potentials, 
it has been suggested that the second virial coefficient 
at the critical temperature $B_{2}(T_{c})$ has a practically constant value
$B_{2}^{c}$, and that short-range tails follow a universal behavior,
provided the second virial coefficient $B_{2}(T)$ is used as the
temperature-like variable instead of the temperature
itself~\cite{vliegenthart00,noro00}. Specifically, it should be possible 
to map the system into a hard-core plus square-well fluid 
with the same hard-core diameter, and a temperature-dependent range
$\delta$ determined so as to give the same $B_{2}$ at the same
reduced temperature $T^*$ as those of the original interaction. As
a consequence, for small enough ranges the reduced
critical temperature should fulfill the relation 
\begin{equation}
\frac{1}{T^{*}_{c}}=\ln\left[1+\frac{1-B_{2}^{c}/B_{2}^{\it HS}}{(1+\delta)^{3}-1}
\right] \, ,
\label{square}
\end{equation}
where $B_{2}^{\it HS}$ is the virial coefficient of the hard-sphere
fluid. Subsequent work~\cite{fu03,roberto06,zhou07,yuka3} has shown 
that, even for short-range interactions, $B_{2}(T_{c})$ changes 
systematically with the attraction range. Nevertheless, at short range 
Eq.~(\ref{square}) does represent quite accurately our results for 
$T^{*}_{c}$. This is shown in Fig.~\ref{fig:tkappa}, where $T^{*}_{c}$ 
as predicted by both 
simulations and theories has been plotted as a function of the
effective square-well range $\delta$, and compared with Eq.~(\ref{square}) 
by taking $B_{2}^{c}$ as a fitting parameter. It can be noticed
that all the theories considered here agree reasonably
well with the simulations, and the discrepancies are smaller
than those found for $\rho^{*}_{c}$. Moreover, for
$\delta\lesssim 0.2$, $T^{*}_{c}$ is indeed well represented by
Eq.~(\ref{square}) with $B_{2}^{c}/B_{2}^{\it HS}\simeq -1.35$. In fact, 
at short range the effective range $\delta$ is much less sensitive than 
$B_{2}(T)$ to small changes in temperature, so that Eq.~(\ref{square}) 
can be satisfied to a very good degree, even though $B_{2}(T_{c})$ is not
actually constant. 

If $B_{2}(T_{c})$ {\em tends} to a finite limit 
$B_{2}^{\it AHS}$ for $\delta\rightarrow 0$, Eq.~(\ref{square}) 
with $B_{2}^{c}=B_{2}^{\it AHS}$ will hold for potentials of arbitrarily short
range down to the AHS limit. Such a behavior was found 
by Largo et al.~\cite{largo2008} for the square-well fluid. We are not able 
to state if this is the case also for the Yukawa fluid since, for the interval 
of $\kappa$ considered in this work, our simulation results do not provide 
a clear evidence of $B_{2}(T_{c})$ saturating to a finite limit.  
In this respect, we should notice that the value 
$B_{2}^{c}/B_{2}^{\it HS}\simeq -1.35$ found here is lower than 
both the result of Largo et al.~\cite{largo2008} $B_{2}^{c}/B_{2}^{HS}=-1.174$ 
and that of Miller and Frenkel~\cite{miller2003} $B_{2}^{c}/B_{2}^{HS}=-1.21$.  

\begin{figure}
\includegraphics[width=8cm]{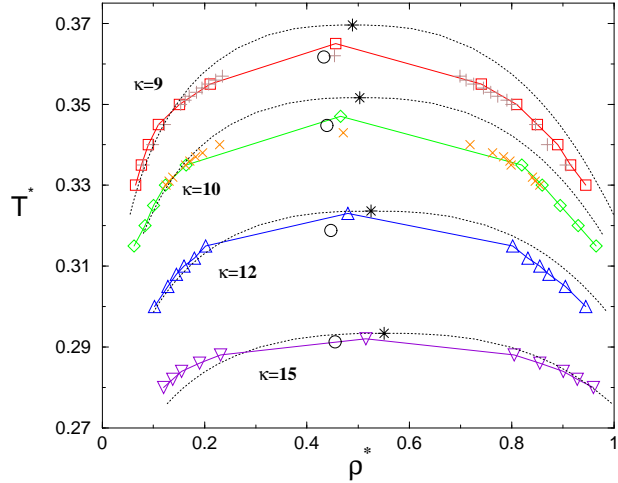}
\caption{Vapor-liquid coexistence curves of SR-HCAY fluid. Open
symbols are NVT-MC data from this work. Other symbols are GC-TMMC
data from Ref.~\cite{singh09}. Dotted lines are S-SCOZA results
from this work. Starbursts and circles are S-SCOZA and M-SCOZA
critical points, respectively.} \label{fig:coex}
\end{figure}

Our new phase equilibrium data of HCAY fluid with $\kappa=9$,
$10$, $12$, and $15$ are shown in Fig.~\ref{fig:coex}, and
compared to those reported by Singh~\cite{singh09} for $\kappa=9$
and $\kappa=10$. An excellent agreement is found between them.
Therefore, both simulation techniques are good for calculating
such properties.
It is worth noting that the NVT-MC
technique is more efficient for the calculation of the coexistence
properties of very short-range systems and at low temperature,
where other simulation techniques such as GEMC or GC-TMMC run into 
considerable difficulties, because the high liquid density of the systems 
makes particle insertion very time-consuming.
Our simulation results for the coexistence curve are reported in
Table~\ref{table2}. 
\begin{table}
\begin{tabular}{ccccccc}
\hline\hline $\hspace{1.0cm}\kappa$\hspace{1.0cm} &\hspace{0.5cm}
$T^{*}$\hspace{0.5cm}  &\hspace{0.5cm} $\rho_{L}$\hspace{0.5cm} &
\hspace{0.5cm}$\rho_{V}$\hspace{0.5cm}   \\

\hline

 9.0    & 0.330  &  0.945     &   0.0650      \\
        & 0.335  &  0.915     &   0.0766      \\
        & 0.340  &  0.890     &   0.0900      \\
        & 0.345  &  0.850     &   0.1100      \\
        & 0.350  &  0.810     &   0.1510      \\
        & 0.355  &  0.741     &   0.2110      \\ \hline

 10.0   & 0.315  &  0.965     &   0.0621      \\
        & 0.320  &  0.930     &   0.0834      \\
        & 0.325  &  0.895     &   0.1005      \\
        & 0.330  &  0.860     &   0.1233      \\
        & 0.335  &  0.820     &   0.1638      \\ \hline

 12.0   & 0.300   &  0.945     &   0.1021      \\
        & 0.305   &  0.905     &   0.1282      \\
        & 0.308   &  0.873     &   0.1454      \\
        & 0.310   &  0.855     &   0.1603      \\
        & 0.312   &  0.832     &   0.1802      \\
        & 0.315   &  0.802     &   0.2022      \\ \hline

 15.0   & 0.280   &  0.960     &   0.1201     \\
        & 0.282   &  0.928     &   0.1380     \\
        & 0.284   &  0.901     &   0.1550     \\
        & 0.286   &  0.855     &   0.1900     \\
        & 0.288   &  0.805     &   0.2310     \\ \hline

\end{tabular}
\caption{Coexistence properties of SR-HCAY fluid at different interaction 
ranges.}
\label{table2}
\end{table}
Figure~\ref{fig:coex} also shows the coexistence curves
and the critical points predicted by S-SCOZA, and the M-SCOZA 
critical points.
M-SCOZA coexistence curves have not been reported, because we
could not rule out the possibility that they may be affected by
significant numerical errors. It appears that the main source of
discrepancy between the simulation and theoretical phase diagrams
lies in the position of the critical points whose densities and
temperatures, as observed above, are both slightly overestimated
by S-SCOZA.\\
\begin{figure}
\includegraphics[width=8cm]{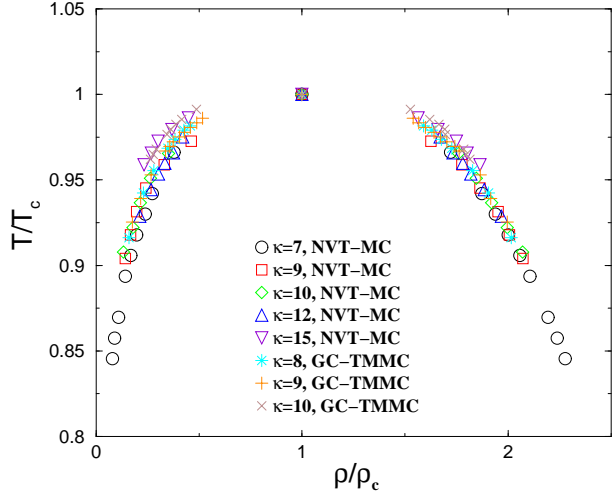}
\caption{Reduced vapor-liquid coexistence curves of
SR-HCAY fluid. NVT-MC data from this work and Ref.~\cite{yuka1}. GC-TMMC data
from Ref.~\cite{singh09}.}
\label{fig:coexscal}
\end{figure}
Figure~\ref{fig:coexscal} shows the reduced density as a function
of the reduced temperature. All curves of short range HCAY fluid
form a single master curve within a very good degree of accuracy,
as it was reported by one of us in a previous work for longer
range~\cite{yuka3}. As the range is decreased, a slight trend
towards a widening of the coexistence curve can be detected, but
this effect remains small for the ranges considered here. Also,
the coexistence curves reported by Singh~\cite{singh09} using
GC-TMMC are on the master curve. The same behavior has been found
for the Mie(n,m) potential~\cite{okumura00, yuri08,pineiro09}. These
results confirm that, even for the short range interactions
considered here, with inverse-range parameter as large as
$\kappa=15$, the vapor-liquid phase diagrams of HCAY fluid obey
the law of corresponding states. This fact could be used to test theoretical
approaches.

\section{Conclusions}
\label{sec:conclusions}

We have carried out a systematic study of the liquid-vapor phase
diagrams for hard-core attractive Yukava potential with $\kappa =
9, 10, 12$ and $15$, using NVT-MC simulation. Our coexistence
curves for $\kappa=9$ and $10$ were compared to those reported by
Singh using GC-TMMC. An excellent agreement was found between
them. Besides, we have shown that reduced coexistence curves of
HCAY fluid from $\kappa=7$ to $\kappa=15$ follow the law of corresponding 
states, as it was shown in a previous work~\cite{yuka3} for longer interaction
ranges. Once again, we have confirmed the linear relationship
between the critical density and inverse critical temperature for
critical simulation data in this interval of $\kappa$. The
simulation results were compared with several theories, including
the standard formulation of SCOZA (S-SCOZA) as well as a modified
version (M-SCOZA) implementing a partial consistency with the
virial route, which is disregarded in the standard version. The
comparison shows that S-SCOZA slightly overestimates both the critical
temperature and the critical density with respect to the
simulation results, while the converse is true for M-SCOZA.
Nevertheless, the agreement with the simulations is quite good for
all the values of $\kappa$ considered in this paper. Both
simulation and theory give a dependence of the critical
temperature as a function of the interaction range in substantial
agreement with Noro-Frenkel scaling. 
However, whether the second virial coefficent at the critical temperature 
$B_{2}(T_{c})$ 
tends to a constant as $\kappa$ diverges, thereby 
making Noro-Frenkel scaling hold at arbitrarily short range as in the
square-well case~\cite{largo2008}, could not be established on the basis 
of our simulation data, and remains to be assessed.   

\section{Acknowledgments}

PO gratefully acknowledges the financial support of the Instituto
Mexicano del Petr\'oleo, under the project D.00476. DP thanks
Alberto Parola, Luciano Reatto and Johan H\o ye for insightful conversation.

\newpage


\begin{thebibliography}{99}
\bibitem{note} In fact, in Ref.~\cite{hoye08} S-SCOZA was mainly intended 
for Lennard-Jones like potentials.
\end{thebibliography}
\end{document}